\documentstyle[aps]{revtex}
\newcommand\be{\begin{equation}}
\newcommand\ee{\end{equation}}
\newcommand\bd{\begin{displaymath}}
\newcommand\ed{\end{displaymath}}
\newcommand\bea{\begin{eqnarray}}
\newcommand\eea{\end{eqnarray}}
\newcommand\nn{\nonumber}
\newcommand\lt{\left}
\newcommand\rt{\right}

\newcommand\bi{\bibitem}

\begin{document}

\title{Superheavy Dark Matter from Thermal Inflation}
\author{
Lam Hui\footnote{Electronic mail: {lhui@fnal.gov}}\,\, and
Ewan D. Stewart\footnote{Electronic mail: {ewand@fnal.gov}} \\
{\em NASA/Fermilab Astrophysics Group} \\
{\em Fermi National Accelerator Laboratory} \\
{\em Box 500, Batavia, IL 60510-0500}
}
\maketitle
\begin{abstract}
It is quite plausible that the mass of the dark matter particle
increases significantly after its freeze-out, due to a 
scalar field rolling to large values. 
We describe a realization of this scenario in the 
context of thermal inflation which naturally gives a cold dark matter
particle with the correct cosmological abundance and a mass around
$10^{10} \, {\rm GeV}$, evading the conventional upper bound of
$10^5 \, {\rm GeV}$.
We also discuss another realization which could produce a
cosmologically interesting abundance of near Planck mass, possibly 
electromagnetically charged, particles.
The detection and observational consequences of superheavy cold dark
matter or WIMPZILLAs are briefly examined.
\end{abstract}

\vspace{-80ex}
{\raggedleft FERMILAB-Pub-98/397-A \\}
\vspace{-\baselineskip}
\vspace{80ex}

\section{Introduction}
\label{intro}

A `model-independent' bound of about $10^5 {\, \rm GeV}$ \cite{griest90}
has often been invoked (see \cite{jungman96} and references therein)
as the largest possible mass the dark matter particle can have.
The derivation of this bound uses the unitarity bound on
the annihilation cross-section, and makes one crucial assumption: 
thermal equilibrium in the early universe.
The unitarity bound on the annihilation cross-section tells us
\be
\langle \sigma_A v \rangle \lesssim \frac{1}{M^2}
\ee
where $M$ is the mass of the dark matter particle,
and $ \langle \sigma_A v \rangle $ is the thermally averaged annihilation
cross-section that appears in the relevant Boltzman equation \cite{early}
(i.e. annihilation per unit time is given by
$ n \langle \sigma_A v \rangle $,
where $n$ is the proper number density of the dark matter particle).
The assumption of thermal equilibrium, on the other hand, tells us that
the freeze-out abundance is given by the thermal distribution
\be
n \sim \lt( M T_f \rt)^\frac{3}{2}
\exp \lt( - \frac{ M }{ T_f } \rt)
\ee
where $T_f$ is the freeze-out temperature.
\footnote{This assumes a cold relic. For a hot relic that freezes out at
$ T_f \gtrsim M $, the number density will be higher leading to a stronger
bound on its mass \cite{early}.}
The exponential suppression implies that $T_f$ cannot be too much smaller
than $M$; hence, combining this with the unitarity bound, it can be
shown that \cite{griest90}
\be\label{bound}
\Omega h^2 \gtrsim 0.1 \lt( \frac{ M }{ 10^5 {\, \rm GeV} } \rt)^2
\ee
where $\Omega$ is the ratio of the mass density in the dark matter particle
to the critical density today, and $h$ is the Hubble constant in units of
$ 100 {\, \rm km \, s^{-1} Mpc^{-1}} $. 
This bound naturally has important implications for dark matter
searches.

Recently, it has been shown that this bound can be evaded by violating the
assumption of thermal equilibrium in the early universe, for instance
by producing the dark matter particle at the end of inflation via
preheating/reheating or gravitational particle-creation \cite{dan,kuzmin98}.
The masses required for a present abundance of $\Omega \sim 1$ are
within a few orders of magnitude of the Hubble parameter at the
end of inflation, i.e. 
$ \sim 10^{12} - 10^{16} {\, \rm GeV} $.
For this reason, they have been called superheavy dark matter or WIMPZILLAs.

Here, we propose a different production mechanism that can also evade the
$10^5 {\, \rm GeV}$ upper bound, and naturally achieves $ \Omega \sim 1 $.
It makes use of a late period of inflation called thermal inflation
\cite{ti,sky,ti2} that occurs at an energy scale of around
$10^6 {\, \rm GeV}$, with a Hubble parameter of the order of
$1 {\, \rm keV}$, which is to be compared with the GUT-scale ordinary
inflation having $ V^{1/4} \sim 10^{16} {\, \rm GeV}$ and
$ H \sim 10^{13} {\, \rm GeV}$ used in \cite{dan,kuzmin98}. 

Thermal inflation provides a natural solution to the Polonyi or moduli
problem \cite{polonyi} that generically arises in string theory
(see also \cite{tinot}).
It occurs when a `flaton', a scalar field with a small
mass and a large vacuum expectation value, is held at the origin
by its finite temperature effective potential.
One gets a few $e$-folds of inflation because the flaton's potential energy
dominates the thermal energy density well before the temperature drops
below the critical temperature for the flaton to start rolling away from
the origin.
The prototypical flaton potential is described in \S \ref{pp}.
Thermal inflation occurs at a very low energy scale which is
the reason it can successfully dilute the potentially harmful moduli
produced after ordinary inflation.
Note that it will also dilute any superheavy dark matter produced at
the end of ordinary 
inflation.

Our idea for dark matter production works roughly as follows.
A particle $\psi$ and its antiparticle $\bar\psi$, which carry some
conserved charge to make them stable, are coupled to the 
flaton.
They are initially massless during the thermal inflation when
the flaton is held at the origin by the finite
temperature effects of $\psi$ and $\bar\psi$. 
After the temperature of the universe drops below the critical
temperature, the 
flaton begins to fast-roll down its potential.
The $\psi$ particle quickly gains mass in the process, which reduces its
annihilation cross-section, and its abundance quickly freezes out.
After that, the flaton continues to roll until it reaches
the true vacuum, acquiring a large expectation value and giving
$\psi$ a large mass.\footnote{A related mechanism has also been
considered in \cite{felder98}.}
The parameters for thermal inflation give a mass for
$\psi$ of around $10^{10} {\, \rm GeV}$,
and work out naturally to give an abundance of
$\Omega_{\psi\bar\psi} \sim 1$.
The details are explained in \S \ref{production}.

This production mechanism evades the
conventional upper bound of $10^5 \, {\rm GeV}$ by giving the particle
$\psi$ a larger annihilation 
cross-section at freeze-out than what one would expect
based on its final mass, and by
entropy production after the freeze-out. 

Since thermal inflation provides a natural solution to the moduli problem,
and since the prediction of $\Omega_{\psi\bar\psi} \sim 1$ follows rather
naturally from its parameters (assuming $\psi$ is stable),
the possibility of a significant fraction of the universe being composed of
superheavy dark matter should be taken seriously.
In \S \ref{detection}, we discuss the observational consequences and
detectability of such dark matter, which could be completely inert,
weakly-interacting, or even electromagnetically or strongly-charged.
Finally, we conclude in \S \ref{conclude} with discussions of
other plausible realizations of the mechanism outlined above,
which include the possibility of producing near Planck mass relics that
could perhaps be electromagnetically charged.

\section{Superheavy dark matter from thermal inflation}
\label{production}

\subsection{Particle physics}
\label{pp}

A simple model of thermal inflation is provided by the superpotential
\cite{susy}
\be\label{Wphi}
W \! \lt( \phi \rt) = \frac{\lambda_\phi \phi^4}{4 M_{\rm Pl}}
\ee
where $\phi$ is the flaton,
and $ M_{\rm Pl} \simeq 2.4 \times 10^{18} {\, \rm GeV} $.
This form for the superpotential can be guaranteed by a $Z_4$ discrete
gauge symmetry because the superpotential is a holomorphic function of
$\phi$, \mbox{i.e.} does not depend on $\phi$'s complex conjugate
$\phi^\dagger$.
The supersymmetric part of the scalar potential is then given by
\be
V_{\rm susy} \! \lt( \phi, \phi^\dagger \rt)
= \lt| \frac{\partial W}{\partial \phi} \rt|^2
= \frac{\lt|\lambda_\phi\rt|^2 \lt|\phi\rt|^6}{M_{\rm Pl}^2}
\ee
In addition, one requires $\phi$'s soft supersymmetry-breaking
mass-squared to be negative.
The scalar potential is then
\be\label{V}
V \! \lt( \phi, \phi^\dagger \rt) = V_{\rm ti} - m_\phi^2 \lt|\phi\rt|^2
+ \lt( \frac{A \lambda_\phi \phi^4}{M_{\rm Pl}} + \mbox{c.c.} \rt)
+ \frac{\lt|\lambda_\phi\rt|^2 \lt|\phi\rt|^6}{M_{\rm Pl}^2}
\ee
where $V_{\rm ti}$ and $A$ are other soft supersymmetry-breaking parameters.
One would expect $ \lt| A \rt| \lesssim m_\phi $.
The scale of the soft supersymmetry-breaking parameters of the
Minimal Supersymmetric Standard Model (MSSM)
is expected to be around $100 \, \mbox{GeV}$ to $1 \, \mbox{TeV}$.
This potential has four degenerate minima
\footnote{The domain walls associated with this degeneracy are most likely
harmless, either because the vacua are identified because of a discrete
gauge symmetry, or because higher order terms, that would generically be
present in the absence of a gauge symmetry, break the degeneracy.}
with $ \lt|\phi\rt| = \phi_{\rm vac} $ where
\be
\phi_{\rm vac}^2 = \frac{ m_\phi M_{\rm Pl} }{ \sqrt{3}\,
\lt|\lambda_\phi\rt| } 
\lt( \sqrt{ 1 + \frac{4\lt|A\rt|^2}{3m_\phi^2} }
 + \frac{2\lt|A\rt|}{\sqrt{3}\, m_\phi} \rt)
\ee
For $ m_\phi = 100 \, \mbox{GeV} $, $ \lt|\lambda_\phi\rt| = 1 $,
and neglecting $A$, this gives $ \phi_{\rm vac} = 10^{10} \, \mbox{GeV} $. 
Requiring zero cosmological constant at the minima gives
\be
V_{\rm ti} = \frac{2}{3} m_\phi^2 \phi_{\rm vac}^2
\lt[ 1 + \frac{\lt|A\rt|}{\sqrt{3}\, m_\phi}
 \lt( \sqrt{ 1 + \frac{4\lt|A\rt|^2}{3m_\phi^2} }
  + \frac{2\lt|A\rt|}{\sqrt{3}\, m_\phi} \rt) \rt]
\label{Vnot}
\ee

Thermal inflation starts when the energy density of the universe starts to
be dominated by $V_{\rm ti}$, and ends a few $e$-folds later when the
temperature drops below that required to hold $\phi$ at $\phi=0$.
$\phi$ then rapidly rolls towards, and oscillates about, the minima of its
potential.
It eventually decays, leaving a radiation dominated universe, at a
temperature $T_{\rm dec}$.
We require
\be\label{TR}
T_{\rm dec} \gtrsim 10 \,\mbox{MeV}
\ee
to avoid interfering with nucleosynthesis. 

In order for $\phi$ to be held at $\phi=0$ by thermal effects during the
thermal inflation, $\phi$ must have unsuppressed interactions with other
fields in the thermal bath.
We will assume these interactions include a coupling of the form
\footnote{
A flaton $\phi$ cannot have a coupling of the form $W \sim \phi^2 \psi$
since it would lead to an unsuppressed quartic self-coupling in the
potential for $\phi$.
Such a coupling can be forbidden by an appropriate gauge symmetry.
The only other possibility would be for $\phi$ to be charged under
some continuous gauge symmetry, which in the vacuum would be broken at a
scale $\sim \phi_{\rm vac} \sim 10^{10} \, \mbox{GeV}$.
We do not consider this possibility here.}
\be\label{Wpsi}
W = \lambda_\psi \phi \bar{\psi} \psi
\ee
with $|\lambda_\psi | \sim 1$.
After thermal inflation, $\psi$ and $\bar{\psi}$ will acquire masses
\be\label{mpsi}
M = \lt| \lambda_\psi \rt| \phi_{\rm vac}
\ee
from this coupling.
For $|\lambda_\psi | = 1$ and $\phi_{\rm vac} = 10^{10} \, \mbox{GeV}$,
this gives $M = 10^{10} \, \mbox{GeV}$.

In order to satisfy the decay constraint, Eq.~(\ref{TR}), we require
$\phi$ to have some, possibly indirect, couplings to the MSSM, beyond
the ever present gravitational couplings.
We will assume this is achieved by having $\psi$ and $\bar{\psi}$ 
charged under SU(3)$\times$SU(2)$\times$U(1).
Other possibilities were considered in Ref.~\cite{sky}.
In order not to interfere with gauge coupling unification,
$\psi$ and $\bar{\psi}$ should form complete representations of SU(5).
These couplings give a decay rate \cite{higgs}
\be
\Gamma_\phi \sim 3 \times 10^{-9} 
\lt( \frac{g_\psi^2 m_\phi^3}{\phi_{\rm vac}^2} \rt)
\ee
where $g_\psi$ is the number of internal degrees of freedom of $\psi$ and
$\bar{\psi}$.
For example, if $\psi = {\bf 5}$ and $\bar{\psi} = {\bf \overline{5}}$
then $g_\psi = 40$.
The decay temperature is therefore \cite{early}
\bea
T_{\rm dec}
& \simeq & g_*^{-\frac{1}{4}} \Gamma_\phi^\frac{1}{2} M_{\rm Pl}^\frac{1}{2}
\sim 2 \times 10^{-5} \lt( \frac{ g_\psi m_\phi^{3/2}
 M_{\rm Pl}^{1/2} }{ \phi_{\rm vac} } \rt) \nn \\
\label{decay}
& \sim & 300 \, \mbox{MeV}
\lt( \frac{ g_\psi }{ 100 } \rt)
\lt( \frac{ m_\phi }{ 100 \, \mbox{GeV} } \rt)^\frac{3}{2}
\lt( \frac{ 10^{10} \, \mbox{GeV} }{ \phi_{\rm vac} } \rt)
\eea
where $g_*$ is the effective number of massless degrees of freedom
in the universe at temperature $T_{\rm dec}$. 
Note that parametric resonance is unlikely to be important because
$\phi$ oscillates around a large vacuum expectation value rather than
the origin. 

Renormalization, amongst other things, will split the degeneracy of
Eq.~(\ref{mpsi}) amongst the various components of $\psi$ and $\bar{\psi}$.
The renormalization will tend to make SU(3) charged components the heaviest, 
and SU(3)$\times$SU(2)$\times$U(1) singlet components, if they exist,
the lightest \cite{dim}.

Our main unjustified assumption is now to assume that $\psi$ and
$\bar{\psi}$ carry opposite charge under some discrete (or continuous)
gauge symmetry under which all other fields are neutral.
This symmetry is, however, helpful in avoiding unwanted superpotential
couplings to MSSM fields.
The lightest component of $\psi$ and $\bar{\psi}$ will then be absolutely
stable and so potentially a dark matter candidate.
\footnote{Throughout this paper, we use the same notation $\psi$ for the
complete representation and 
its lightest component (the dark matter).}
For example, we could have a $Z_8$ discrete gauge symmetry, under which
$\phi$, $\psi$, and $\bar{\psi}$ have charges $2$, $-1$, and $-1$,
respectively.
This would guarantee Eqs.~(\ref{Wphi}) and~(\ref{Wpsi}),
and after $\phi$ acquires its vacuum expectation value,
the $Z_8$ will be broken down to a $Z_2$
under which only $\psi$ and $\bar{\psi}$ are charged.
A simple choice of representations that satisfies the discrete
anomaly cancellation conditions \cite{anomaly} is
$ \psi = {\bf 16} + {\bf 1} $ and
$ \bar{\psi} = {\bf \overline{16}} + {\bf 1} $
of SO(10).

\subsection{Abundance}

During thermal inflation, $\psi$ and $\bar{\psi}$ will be relativistic and
in thermal equilibrium.
Their number density will therefore be
\be
n \! \lt( T \rt) = \frac{7 \zeta(3) g_\psi T^3}{8 \pi^2}
\label{thermal}
\ee
where $\zeta(3) \simeq 1.202$, and $g_\psi$ is the number of internal
degrees of 
freedom of $\psi$ and $\bar{\psi}$.
\footnote{We have assumed, as is appropriate for a supersymmetric theory,
that there are equal numbers of bosonic and fermionic degrees of freedom
($7/8=(1+3/4)/2$).}
If $ \psi = {\bf 16} + {\bf 1} $ and
$ \bar{\psi} = {\bf \overline{16}} + {\bf 1} $
of SO(10), as in the example of the previous section, then
$ g_\psi = 136 $.
Through the coupling of Eq.~(\ref{Wpsi}), $\psi$ and $\bar{\psi}$ will
generate the finite temperature effective potential for $\phi$
\be
V_T \! \lt( \phi \rt) = V_{\rm ti}
+ \lt( \frac{g_\psi}{32} \lt|\lambda_\psi\rt|^2 T^2 - m_\phi^2 \rt)
\lt|\phi\rt|^2 + \ldots
\ee
Thermal inflation will therefore end at the temperature
\be\label{Tc}
T_{\rm c} = \frac{4\sqrt{2}\, m_\phi}{\sqrt{g_\psi}\, \lt|\lambda_\psi\rt|}
\ee
when $\phi$ begins to roll away from $\phi=0$.
Shortly afterwards, the abundance of $\psi$ and $\bar{\psi}$ freezes
out. Meanwhile, $\phi$ will continue to roll towards its 
vacuum expectation
value $\lt|\phi\rt| = \phi_{\rm vac}$.
Once $\phi$ acquires its vacuum expectation value $\lt|\phi\rt| =
\phi_{\rm vac}$, 
the coupling of Eq.~(\ref{Wpsi}) will give $\psi$ and $\bar{\psi}$ masses
\be
M = \lt| \lambda_\psi \rt| \phi_{\rm vac}
\ee

The freeze-out abundance of $\psi$ and $\bar{\psi}$ can be estimated
as follows.
First, it is important to keep in mind that the temperature does not drop
as $\phi$ rolls from $0$ to $\phi_{\rm vac}$.
This is because the time-scale for the roll-over is $m_\phi^{-1}$,
which is much smaller than the Hubble time
$H_{\rm ti}^{-1} \sim m_\phi^{-1} \phi_{\rm vac}^{-1} M_{\rm pl} \gg
m_\phi^{-1}$. 
Hence $T = T_{\rm c}$ throughout the roll-over.
The freeze-out occurs as $\psi$ gains mass and begins to
become non-relativistic when its thermal abundance is given by
\be\label{nM}
n(m_\psi) = g_\psi \lt( \frac{m_\psi T_{\rm c}}{2\pi} \rt)^\frac{3}{2}
\exp \lt( - \frac{m_\psi}{T_{\rm c}} \rt)
\ee
where $m_\psi$ denotes the $\phi$ dependent mass of $\psi$, 
$m_\psi = \lt| \lambda_\psi \phi \rt|$, which increases as $\phi$
rolls down the potential. 
In other words, in contrast with the usual freeze-out calculation,
it is $m_\psi$ that is changing with time rather than the temperature.
The freeze-out abundance is determined by equating the
annihilation rate $\Gamma_{\psi\bar{\psi}}$ with the
inverse-time-scale of the problem at hand, \mbox{i.e.} $m_\phi$.
\be
\Gamma_{\psi\bar{\psi}} \! \lt( m_\psi \rt) \sim n \! \lt( m_\psi \rt)
\langle \sigma \lt| v \rt| \rangle \! \lt( m_\psi \rt) \sim m_\phi
\ee
Using the fact that $\langle \sigma \lt| v \rt| \rangle \lesssim 1/{m_\psi^2}$,
and that $m_\phi \sim \lt|\lambda_\psi \rt| T_{\rm c}$, it is not hard to
see that the freeze-out occurs when $m_\psi \sim T_{\rm c}$, and that the
freeze-out abundance is given by Eq.~(\ref{thermal}) with $T = T_{\rm c}$,
suppressed by at most a factor $\lt| \lambda_\psi \rt|$. So
for $\lt| \lambda_\psi \rt| \sim 1$, and to within an order of
magnitude, there is no significant net annihilation of $\psi$ and $\bar\psi$
after the beginning of the roll-over,
and the freeze-out occurs well before $\phi$ reaches $\phi_{\rm vac}$.

After the freeze-out, $\phi$ will continue to roll towards its vacuum
expectation value, increasing the mass of $\psi$ to a large value.
It can be checked that the annihilation rate by the time $\phi$ reaches
the minimum is negligible
\footnote{If $\lt| \lambda_\psi \rt|$ is small, the fifth power of
$\lt| \lambda_\psi \rt|$ in this formula could make
$\Gamma_{\psi\bar{\psi}}$ significant which means that there would
be some annihilation of $\psi$ and $\bar\psi$.
However, in this case the cosmological abundance will also be boosted
up by a higher $T_c$ (see Eqs. (\ref{Tc}) and~(\ref{Omega})).} 
\be
\Gamma_{\psi\bar{\psi}} \lesssim \frac{T_{\rm c}^3}{M^2}
\sim \frac{m_\phi^3}{\lt| \lambda_\psi \rt|^5 \phi_{\rm vac}^2}
\sim \lt( \frac{\lt| \lambda_\phi \rt|^{3/2} m_\phi^{1/2}}
{\lt| \lambda_\psi \rt|^5 M_{\rm Pl}^{1/2}} \rt) H_{\rm ti} \ll H_{\rm ti}
\ee
Subsequently, $\phi$ will oscillate around $\phi_{\rm vac}$.
\footnote{Note that the backreaction of the finite density of $\psi$
particles on $\phi$'s potential is negligible.}

One might worry that during the oscillation,
a significant amount of annihilation or production of $\psi$ and
$\bar\psi$ might occur if $\phi$ returns to small values.
The rapid build up of gradient energy will prevent $\phi$ from
returning to small values except in a few isolated places.
In addition, $\phi$ would eventually be prevented from returning to
small values by the Hubble expansion.    
Strings with walls attached are also formed, which will likely
disappear quickly \cite{shellard}. Their direct radiation into the
heavy $\psi$ particles is heavily suppressed because $M_\psi \gg
m_\phi$. On the other hand, the $\psi$ particles are light in the
cores of the strings, and 
could be created and trapped there. If a string loop carries a net
$\psi$-charge, it will be 
released in the form of (heavy) $\psi$ particles when the loop annihilates.
If we assume each string produces of the order of one $\psi$ particle,
our rough estimate of the $\psi$ abundance should still be valid.

The energy density in $\psi$ and $\bar{\psi}$ will then scale with that
of the oscillating flaton until the flaton finally decays, leaving a
radiation dominated universe at a temperature $T_{\rm dec}$. 
The energy density of $\psi$ and $\bar{\psi}$ will then scale with the
entropy density of the universe, $s$.
The current value of the entropy density is
\be
s_0 = 2.2 \times 10^{-38} \, \mbox{GeV}^3
\ee
Finally, we wish to compare the current energy density of $\psi$ and
$\bar{\psi}$ with the critical density
\be
3 H_0^2 = 3 \times 10^{-47} \, \mbox{GeV}^4
\ee
For $\psi$ and $\bar{\psi}$ to be a viable dark matter candidate we
require
\be
\Omega_{\psi\bar{\psi}} \equiv \frac{\rho_{\psi\bar{\psi}}}{3 H_0^2}
\sim 0.3
\ee
Putting everything together we get
\be
\Omega_{\psi\bar{\psi}} =
n \! \lt( T_{\rm c} \rt) M
\lt( \frac{\rho_{\rm dec}}{V_{\rm ti}} \rt) 
\lt( \frac{s_0}{s_{\rm dec}} \rt) 
\lt( \frac{1}{3 H_0^2} \rt)
\ee
where $\rho_{\rm dec}$ is the energy density of the universe at the end of
the flaton decay.
Therefore, using $ \rho_{\rm dec} = \frac{3}{4} T_{\rm dec} s_{\rm dec} $,
\bd
\Omega_{\psi\bar{\psi}} =
\lt| \lambda_\phi \rt| \lt|\lambda_\psi\rt|^{-2}
\lt( \frac{ g_\psi }{ 100 } \rt)^\frac{1}{2}
\lt( \frac{ m_\phi }{ 100 \, \mbox{GeV} } \rt)^\frac{3}{2}
\lt( \frac{ 5 \times 10^4 \phi_{\rm vac} T_{\rm dec} }
 { g_\psi m_\phi^{3/2} M_{\rm Pl}^{1/2} } \rt)
\ed
\be\label{Omega}
\times
\lt( \frac{ 8 \pi^2 n \! \lt( T_{\rm c} \rt) }
 { 7 \zeta(3) g_\psi T_{\rm c}^3 } \rt)
\lt( \frac{ \sqrt{g_\psi}\, \lt| \lambda_\psi \rt| T_{\rm c} }
 { 4 \sqrt{2}\, m_\phi } \rt)^3
\lt( \frac{ M }{ \lt| \lambda_\psi \rt| \phi_{\rm vac} } \rt)
\lt( \frac{ 2 m_\phi^2 \phi_{\rm vac}^2 }{ 3 V_{\rm ti} } \rt)
\lt( \frac{ m_\phi M_{\rm Pl} }
 { \sqrt{3}\, \lt|\lambda_\phi\rt| \phi_{\rm vac}^2} \rt)
\ee
where all factors in brackets are of order $1$. 
We have displayed explicitly the assumed relations between
the various quantities,
\mbox{e.g.} $M = |\lambda_\psi | \phi_{\rm vac}$, etc.

\section{Observational Constraints and Detection}
\label{detection}

The above simple model leaves open the question of what kind of
interaction $\psi$ has with ordinary matter.
The same is true of other production mechanisms of superheavy dark
matter \cite{dan,kuzmin98}. Let us go through the different
possibilities one by one.

{\it Electromagnetically charged.} These have been referred to as CHAMPs in
the literature \cite{champs}. At late times, they primarily take the form of
$p^+ \psi^-$ (hydrogen with a heavy ``electron'' which has very low
cross-section with other atoms), $\psi^+ e^-$ (heavy hydrogen) or 
$\psi^- {\rm He}^{++} e^-$ ($\psi^-$ bound to the helium nucleus to
make another kind of heavy hydrogen). Various constraints exist on
such particles, ranging from the absence of
heavy-hydrogen-like atoms in water to nondetection in $\gamma$-ray and
cosmic-ray detectors \cite{champs}. By far, the strongest constraint appears
to come from the existence of old neutron stars \cite{gould90}, where
only $M \gtrsim 10^{16} {\, \rm GeV}$ is allowed.
Otherwise, a sufficient net number of $\psi^+$ particles collects in
the neutron 
star, forms a black hole in the center and eats up the star on a short
time scale; this is in part because the 
hydrogen-heavy-hydrogen scattering cross-section is high, given by the
square of the Bohr radius in the low velocity limit $\sigma \sim 10^{-17} {\,
\rm cm^{-2}}$. 
However, if the abundance of $\psi^+$ and that of $\psi^-$ bound to helium
are the same in the halo, the constraint is weakened to
$M \gtrsim 10^{10} {\, \rm GeV}$. The conventional bound of 
$M \lesssim 10^5 {\, \rm GeV}$ would then be fatal to the existence of
such particles. The kind of 
production mechanism like the one proposed here, or elsewhere, which
evades the conventional bound, could resurrect the
intriguing idea that the dark matter can be charged.
But as we can see, significant astrophysical constraints already
exist. It should be noted that the near Planck mass relic that will be
discussed in \S \ref{conclude} satisfies even the demanding bound of
$M \gtrsim 10^{16} {\, \rm GeV}$. The economic importance of such
stable massive electromagnetically charged particles cannot be
overestimated. For example, they could be used to catalyze nuclear
fusion \cite{champs}.

{\it Strongly charged.} These have been referred to as SIMPs in
the literature \cite{starkman90,simps}. Significant bounds on their
masses, if they have 
significant cosmological abundances, come from
nucleosynthesis as well as the absence of anomalously heavy isotopes
of familiar nuclei \cite{mohapatra}. A
systematic study of constraints from direct detection and the
existence of old neutron stars and the Earth was made in Ref.
\cite{starkman90}. Assuming 
there is no $\psi$-$\bar\psi$ asymmetry, 
the neutron-star argument and underground plus balloon experiments
provide competitive bounds: only $M
\gtrsim 10^{8} - 10^{10} {\, \rm GeV}$ is allowed for a
$\psi$-proton 
cross-section of $\sigma \sim 10^{-30} - 10^{-25} {\, \rm cm^{-2}}$.
As in the case of electromagnetically charged dark matter,
the possibility of producing them without overclosing the universe
gives such dark matter candidates a new life.

{\it Weakly charged.} Naturally, no significant constraints exist
if $\psi$ has only weak-scale interactions like the neutralino
(i.e. $\sigma \sim 10^{-44} (m_n/{\, \rm
GeV})^4 {\, \rm cm^{-2}}$ 
in the large $M$ limit,
where $m_n$ is the mass of the relevant nucleon).
Direct detection appears rather difficult simply because the
halo number density scales as $M^{-1}$, and
the neutralino with mass $\sim 100 {\, \rm GeV}$ is already
difficult to detect. Note that a large mass does not
increase significantly the nuclear recoil: $\Delta E \sim
M^2 m_n v^2 / (M + m_n)^2$ where $v \sim 200 {\, \rm km \, s^{-1}}$ is
the average halo velocity of these particles;
in the large $M$ limit, $\Delta E$ is asymptotically
$M$-independent.
Indirect detection might seem even more hopeless. Not only does the halo
number density drop by a factor of $M$, the
annihilation (which gives rise to neutrinos) rate is suppressed by
$M^2$ according to the unitary bound. However, 
three opposing factors help us here. First, on a sufficiently long time scale,
the neutrino flux from $\psi$-$\bar\psi$ annihilation in the core of
the Sun or the Earth is determined not by the annihilation rate, but
by the capture rate, which depends on the scattering cross-section
with nucleons and is not heavily suppressed. Second, indirect
detection works by observing muons that result from the interaction of
the neutrinos with the rocks of the Earth. The cross-section for
producing muons and the range of the muons both scale up with energy,
and hence the mass of $\psi$. A detailed calculation taking into account
these effects as well as other relevant ones will
be presented in a forthcoming paper.

{\it Completely neutral.}
$\psi$ could be a singlet under SU(3)$\times$SU(2)$\times$U(1).
It is of course virtually impossible to detect such particles,
except by their gravitational effects.

Lastly, superheavy dark matter has been postulated as the origin of
ultra-high-energy cosmic-ray events at energies $\gtrsim 10^{10} {\,
\rm GeV}$, above the Greisen-Zatsepin-Kuzmin cut-off
\cite{berez97,birkel98,dan,kuzmin98}. The parameters of the model
presented in the last
section are sufficiently flexible to allow a mass of $M \sim
10^{11} {\, \rm GeV}$ to explain such events.
The $\psi$ particles cannot by themselves be the primaries
because of the large mass, even if they have hadronic interactions 
\cite{albuquerque98,mohapatra}. The simplest way is to have them decay into
hadrons, e.g. protons, which reach the Earth's atmosphere.
But then, one has to invoke special reasons to explain why they are
not stable but sufficiently long-lived. 

\section{Discussion}
\label{conclude}

We have shown that a simple and well-motivated model of thermal
inflation naturally produces dark matter particles $\psi$ and 
$\bar\psi$ of mass $ M \sim 10^{10} {\, \rm GeV} $ with a cosmological
abundance in the correct range. 
As our mechanism is implemented by thermal inflation, which solves the
moduli problem by late entropy production, it is robust against
such dilution.
%It is uninteresting to note that unlike other mechanisms
%\cite{dan,kuzmin98}, our mechanism produces a particle of mass $M$
%well above the Hubble parameter at the time of production. 

The same general mechanism could also be applied to what one might call
`moduli thermal inflation' to produce near Planck mass particles at low
energy scales.
Moduli thermal inflation is a limit of thermal inflation in which the
flaton is replaced by a modulus, so that roughly speaking
$ \lt| \lambda_\phi \rt| \sim m_\phi / M_{\rm Pl} $ in Eq.~(\ref{V}).
$\phi_{\rm vac}$ will then be of the order of the Planck scale.
In the context of string theory, it is then very reasonable to assume that
the vacuum expectation value of the modulus, $ \phi = \phi_{\rm vac} $,
corresponds to another `origin' in field space where new fields become
light; for example one could have superpotential couplings of the form
$ W = \lambda_\chi \lt( \phi - \phi_{\rm vac} \rt) \bar{\chi} \chi $.
Such a `coupled' modulus would appear like an ordinary scalar field
(\mbox{e.g.} a squark or slepton field) in the true vacuum.
\footnote{It would be an excellent candidate for an Affleck-Dine field
\cite{AD}.}
The decay temperature would then no longer scale as in Eq.~(\ref{decay})
but instead could be as high as $V_{\rm ti}^{1/4}$.
Putting these modifications into Eq.~(\ref{Omega}) would also give us
a value of $\Omega_{\bar\psi\psi}$ in the neighborhood of 1.
However, another consequence of having $ \phi_{\rm vac} \sim M_{\rm Pl} $
is that one could get a significant number of $e$-folds of non-slow-roll
inflation as $\phi$ rolls from $\phi \sim 0 $ to
$ \phi \sim \phi_{\rm vac} $.
To avoid this inflation diluting the $\psi$ particles too much,
one would require
$ m_\phi \gtrsim 10 V_{\rm ti}^{1/2} / M_{\rm Pl} $
and so $ \phi_{\rm vac} \lesssim M_{\rm Pl} / 10 $.
This would limit the final mass of the $\psi$ particles to be
$ M \lesssim \mbox{few} \times 10^{17} {\, \rm GeV} $.
However, this is still above even the stringent
limit on electromagnetically charged dark matter obtained in
Ref.~\cite{gould90}. 

Note that because moduli thermal inflation occurs at too high
an energy scale to solve the moduli problem, this scenario
would only be viable if there were no moduli
problem \footnote{
For example, because all the moduli are of this coupled type,
rather than the decoupled type that give rise to the moduli problem.
How one fits the dilaton into such a picture is unclear though.}
because otherwise the $\psi$ particles would be diluted by
another epoch of thermal inflation, or some other late entropy
production, that would be needed to dilute the decoupled moduli produced
at the end of the moduli thermal inflation.

A related scenario could emerge from some of the more plausible models
of inflation \cite{lisa}.
Here $\psi$ or $\bar\psi$ is the inflaton, which holds $\phi$ at zero
by the hybrid inflation mechanism \cite{hybrid} rather than thermal
effects.
One could then get dark matter in the form of charged, near Planck mass,
inflatons!

\subsection*{Acknowledgements}
We thank Rocky Kolb, Dan Chung, Josh Frieman, and Andrew Sornborger for useful
discussions. 
LH thanks the German-American Young Scholars' Institute on 
Astroparticle Physics and the Aspen Center for Physics for hospitality.
This work was supported by the DOE and
the NASA grant NAG 5-7092 at Fermilab.

\end{document}